\documentstyle[aps,twocolumn,prl,epsf,amssymb,floats]{revtex} 
\begin{document}

\twocolumn[    
\hsize\textwidth\columnwidth\hsize\csname @twocolumnfalse\endcsname    

\title{Phase separation due to quantum mechanical correlations}
\author{James K. Freericks$^*$, Elliott H. Lieb$^{\dagger}$, and
Daniel Ueltschi$^\ddagger$}
\address{$^*$Department of Physics, Georgetown University, Washington, DC
20057, USA\\
$^\dagger$Departments of Mathematics and Physics, Princeton University, Princeton, NJ 08544, 
USA\\
$^\ddagger$Department of Mathematics, University of California, Davis, CA 95616,
USA}

\date{\today}
\maketitle

\widetext
\begin{abstract}
Can phase separation be induced by strong electron correlations?  We
present a theorem that affirmatively answers this question in the
Falicov-Kimball model away from half-filling, for any dimension.  In the ground
state the itinerant electrons are spatially separated from
the classical particles.
\end{abstract}

]      

\narrowtext
The Falicov-Kimball (FK) model~\cite{FK69}
can be viewed as a modification
of the Hubbard (H) model~\cite{H63} in which one species of electrons (say
spin down) has infinite mass. As such, its relation to the latter is
similar to the relation of the Ising model to the quantum Heisenberg
model~\cite{KL86}.  Alternatively, it can be viewed as a model
of itinerant electrons and immobile ions. It possesses long range order at
low temperature in two or more dimensions at half filling, and this
checkerboard state (and higher-period 
generalizations~\cite{GJL92,K94,K98,HK01}) remain to-date as the 
only examples of crystallization
into a perfectly ordered structure whose periodicity is not that of the
underlying lattice.

In this paper, we report a theorem on the existence of phase separation 
in the ground state of the FK model, away from half-filling and for large 
repulsion between the particles. `Phase separation', or `segregation', means 
that the system splits into two large domains, one being
occupied by the classical particles, and the other by the quantum particles. 
In the language of the H model, this would mean segregation of spin up particles
from spin down particles --- resulting in a ferromagnetic state. 
The question of whether strong interactions can drive quantum (electronic)
systems to phase separate was posed over ten years ago for the FK model
\cite{FF90} and the H model \cite{KE90}.  This work
is a rigorous proof of this long-standing conjecture for the FK model.
Further discussion of the
relation between the FK and the H model is given later.

Phase separation is not new in classical
lattice models. It is present in the Ising model, and
in many other classical models. It also occurs in the FK model at half-filling
for some densities, as was proved in \cite{K98}. In these examples it is mainly a {\it
local} phenomenon --- interactions (or `effective interactions' in the case of
FK) tend to dislike boundaries, and the state with minimum boundary is phase
separated.

Away from half-filling, the
electrons of the FK model are in delocalized wave functions, and their energy cannot be
written as a sum of local terms. While we prove that the ground state energy is roughly
proportional to the boundary between occupied and empty sites, the mechanism is {\it
nonlocal} and genuinely quantum mechanical.

The FK Hamiltonian~\cite{FK69} is
\begin{equation}
H=-\sum_{{\bf x},{\bf y}\in {\Omega}} t({\bf x} - {\bf y}) c^{\dagger}_{\bf x}c_{\bf y}+
U\sum_{{\bf x}\in {\Omega}}
c^{\dagger}_{\bf x}c_{\bf x}w_{\bf x}.
\end{equation}
Here, $t({\bf x}-{\bf y})$ is the hopping coefficient between sites ${\bf x}$ and
${\bf y}$; it is translation
invariant, but may depend on the direction (this allows consideration of general Bravais
lattices). $c^{\dagger}_{\bf x}$
and $c_{\bf x}$ are creation and annihilation operators
for a spinless electron at site ${\bf x}$, and $w_{\bf x} = 1$ or 0 is a classical 
variable
that denotes the presence  or the absence of an ion at ${\bf x}$. (Spin degrees of freedom
have trivial behavior and are left aside here.)
$\Omega \subset {\mathbb Z}^d$ is 
a finite $d$-dimensional lattice, and
$U$ is the on-site repulsion between the two species of particles.  For any 
given configuration $w = \{w_{\bf x}\}$ of
classical particles, the ground state for $N_{\rm e}$ electrons is determined by
diagonalizing a one-body operator given by the above Hamiltonian, and
filling in the lowest $N_{\rm e}$ states.  The main question is to find which 
configuration $w$,
with a given number of classical particles $N_{\rm c} = \sum_{\bf x} w_{\bf x}$,
minimizes the energy of the electrons.

Our theorem states upper and lower bounds for the energy of $N_{\rm e}$ 
electrons, for a given configuration $w$ of the classical particles. For 
orientation, let us consider first
a configuration where the sites devoid of classical particles form a large, 
`compact' region. The
expected energy of the electrons is a bulk
term that scales like the volume of this region, and a correction that 
scales like its boundary. Our main result is a proof of this conjecture for
{\it all} configurations, not only `nice' ones.

We need some notation.
Let $\Lambda = \{ {\bf x} \in \Omega: w_{\bf x} = 0 \}$ 
denote the set of empty sites for the configuration $w$, and $\partial\Lambda$ 
its boundary, $\partial\Lambda = \{ {\bf x}\in\Lambda, {\rm dist}({\bf x},\Omega\setminus
\Lambda)=1 \}$. Their respective
number of sites are $|\Lambda|$ and $|\partial\Lambda|$. We
write $E(N_{\rm e},w)$ for the ground
state energy of $N_{\rm e}$ electrons in the configuration $w$.
An important quantity is $n=n_{\rm e}/(1-n_{\rm c})$, where 
$n_{\rm e} = N_{\rm e}/|\Omega|$ and $n_{\rm c} = \sum_{\bf x} w_{\bf x}/|\Omega|$
are the densities for quantum and classical particles respectively. 
It represents the
electronic density that would exist inside $\Lambda$, if all electrons
live inside the domain devoid of classical particles.
Let $e(n)$ be the usual kinetic energy per site for noninteracting electrons with
density $n$ in the thermodynamic limit (its expression is recalled below, see Eq.\
(\ref{eq: e_fermi})).

\vspace{1mm}
\noindent
{\bf Theorem.} {\it For all $\Lambda$ we have 
upper and lower bounds,
$e(n) |\Lambda| + \alpha'(n) |\partial\Lambda| \ge E(N_e,w) \ge e(n) |\Lambda|
+ \alpha(n,U) |\partial\Lambda|$.

Here, $\alpha' (n)$  and $\alpha (n, U)$ are explicitly given positive
functions. For nearest-neighbor hoppings (i.e., $t({\bf x}) \neq 0$ if $|{\bf x}|=1$, $t({\bf
x})=0$ otherwise), $\alpha(n,U) = \alpha(n) -
\gamma(U)$, where $\alpha(n) = \alpha(1-n)$ is strictly positive for 
$0<n<1$, and $\gamma(U)$ satisfies $\lim_{U\to\infty} U \gamma(U) = 8d^2$.
}
\vspace{1mm}

The theorem clearly leads to segregation because low-energy configurations must
have a small boundary, i.e.\ there is a relatively small number of empty sites that are 
neighbors of the classical particles. Furthermore, due to the large repulsion, 
electrons are essentially located in the empty domain. The boundary of the
configuration of classical particles that minimizes the energy is smaller than the minimum
possible boundary times the ratio between the upper and lower bounds
(which is a large constant).

Phase separation in the FK model for large $U$ was conjectured 
in~\cite{FF90} and is in stark contrast to the situation at
$n_{\rm c} = n_{\rm e} = \frac12$, where long-range order of checkerboard type occurs, as was 
established in~\cite{KL86}. The theorem is proved in~\cite{FLU01}, and 
extends results for $d=1$~\cite{L92} and 
$d=\infty$~\cite{FGM99} to all dimensions, in particular to the dimensions
2 and 3 that are of great physical relevance. Its proof relies
on a result of Li and Yau for the Laplace operator in the 
continuum~\cite{LY83} (see also~\cite{LL01}, Theorem 12.3). We explain 
this elegant proof below in the case of a lattice,
thereby proving the lower bound for $U=\infty$
without the boundary correction. To include 
this boundary correction requires considerably more effort, and we 
refer to~\cite{FLU01} for further details.

\noindent{\it Proof with $\alpha(n)=0$ and $U=\infty$:}
The ground-state energy $e(n)$ for noninteracting electrons with density
$n$ is found in the usual fashion: (i) define a Fermi energy 
$\varepsilon_{\rm F}$ via
\begin{equation}
\frac{1}{(2\pi)^d} \int_{\epsilon({\bf k})\le\varepsilon_{\rm F}} {\rm d}^d k = n,
\end{equation}
with the band structure $\epsilon({\bf k})=-\sum_{\bf x} t({\bf x}) \exp
(-{\rm i} {\bf k\cdot x})$ and (ii) do the integration
\begin{equation}
\frac{1}{(2\pi)^d} \int_{\epsilon({\bf k})\le\varepsilon_{\rm F}} \epsilon({\bf k})
\, {\rm d}^d k = e(n).
\label{eq: e_fermi}
\end{equation}

The electrons are forbidden to lie on any site
occupied by the classical particles.  Then the eigenfunctions for a given configuration
$w$ are found by diagonalizing the projection of the hopping matrix
onto $\Lambda$.   Let
$\phi_\beta({\bf y};w)$ denote the orthonormal eigenvectors
of the Hamiltonian
indexed by $\beta=1,..., |\Omega|-N_{\rm c}$ for the configuration
$w$ with eigenvalues $\epsilon_\beta(w)$.
We choose the ordering of the labels such that $\epsilon_1(w)\le
\epsilon_2(w)\le ... \le \epsilon_{|\Omega|-N_{\rm c}}(w)$. We
also extend the definition of the eigenvectors to all of ${\mathbb Z}^d$
setting $\phi_{\beta}({\bf y};w)=0$ for all ${\bf y}\notin \Omega$.
The ground state energy for $N_{\rm e}$ electrons and $N_{\rm c}$ classical particles in the
configuration $w$ is
\begin{equation}
E(N_{\rm e},w)=\sum_{\beta=1}^{N_{\rm e}}
\epsilon_\beta(w),
\label{eq: gsdef}
\end{equation}
and the ground-state energy $E_{\rm g.s.}(N_{\rm e}, N_{\rm c})$
is the minimum of Eq.~(\ref{eq: gsdef})
over all configurations that contain $N_{\rm c}$ classical particles.

Using the definition of the eigenvectors allows us to write the ground
state energy as 
\begin{equation}
E(N_{\rm e},w) = \sum_{\beta=1}^{N_{\rm e}}
\sum_{{\bf y},{\bf z}\in {\mathbb Z}^d}\phi_\beta^*({\bf y};w)[-t({\bf y} - {\bf z})]
\phi_\beta({\bf z};w).
\label{eq: gs2}
\end{equation}
Now, we define the Fourier transform of the eigenfunctions by
\begin{equation}
f_{\beta}({\bf k};w)=\sum_{{\bf y}\in {\mathbb Z}^d} {\rm e}^{{\rm i}{\bf k}\cdot{\bf y}}
\phi_\beta({\bf y};w),
\label{eq: ft}
\end{equation}
for every wave vector {\bf k} in the Brillouin zone $[0,2\pi]^d$. It is well-known
that the energy can be expressed as
\begin{equation}
E(N_{\rm e},w)=\frac{1}{(2\pi)^d}
\int {\rm d}^d k \, \epsilon({\bf k})\rho ({\bf k};w);
\label{eq: gsmin}
\end{equation}
here and in the following the integral is over $[0,2\pi]^d$ and we introduced
the density function
\begin{equation}
\rho({\bf k};w) = \sum_{\beta=1}^{N_{\rm e}} |f_{\beta}({\bf k};w)|^2.
\end{equation}
The density function is obviously positive and is bounded above by $|\Lambda|$.
Indeed, we can write
\begin{equation}
\rho({\bf k};w) = \sum_{\beta=1}^{N_{\rm e}} \sum_{{\bf y},{\bf z}\in {\mathbb Z}^d} 
{\rm e}^{-{\rm i} {\bf k}\cdot{\bf z}} \phi_\beta^*({z};w)
{\rm e}^{{\rm i} {\bf k}\cdot{\bf y}} \phi_\beta({y};w);
\end{equation}
this can be rewritten as
\begin{equation}
\rho({\bf k};w) = \sum_{{\bf y},{\bf z}\in \Lambda} {\rm e}^{-{\rm i} {\bf k\cdot{\bf z}}}
\rho({\bf z}, {\bf y}; w) {\rm e}^{{\rm i} {\bf k\cdot{\bf y}}}
\end{equation}
with $\rho({\bf z}, {\bf y}; w) = \sum_{\beta=1}^{N_{\rm e}} \phi_\beta^*({\bf z};w)
\phi_\beta({\bf y};w)$.  The matrix $\rho$ is a positive semidefinite 
matrix bounded by 1 (it is the projector onto the lowest $N_{\rm e}$ eigenvectors).  Hence, 
\begin{equation}
\rho({\bf k};w) \le \sum_{{\bf y}\in \Lambda}| {\rm e}^{{\rm i} {\bf k\cdot{\bf y}}}|^2
= |\Lambda|.
\end{equation}
Furthermore, the density function satisfies a sum rule
\begin{equation}
\frac{1}{(2\pi)^d} \int {\rm d}^d k \, \rho({\bf k};w) =
\sum_{{\bf y} \in {\Lambda}} \rho({\bf y},{\bf y};w) = N_{\rm e}.
\end{equation}
One gets a lower bound for $E(N_{\rm e}, w)$ by minimizing the right side of
(\ref{eq: gsmin}) over all functions $\rho$ satisfying
$0\leq\rho\leq|\Lambda|$ and whose integral is $N_{\rm e}$. This is
the `bathtub principle', see~\cite{LL01}, Theorem 1.14; the minimizer is $\rho({\bf k})=|\Lambda|$
for $\{ {\bf k}:\epsilon({\bf k})\le \varepsilon_{\rm F}\}$ and $\rho({\bf k})=0$
otherwise, with $\varepsilon_{\rm F}$ defined by Eq.~(\ref{eq: e_fermi}). This 
amounts to filling the
lowest eigenvalues of the infinite-lattice.

The proof of the upper bound for the energy proceeds by
forming an `average' Hamiltonian by translating 
and rotating the original configuration $w$ over a large but finite 
subset of ${\mathbb Z}^d$ (with periodic boundary conditions).  Then on
a bipartite lattice, one can show by concavity of the sum of the lowest
$N$ eigenvalues of a matrix, that the averaged Hamiltonian provides an upper
bound to the ground state energy.  But the magnitude of the averaged hopping is
determined by the size of the boundary, which eventually yields the desired upper
bound.  Extending the proof above to provide the lower bound is much more 
complicated and relies on a detailed technical examination of the influence
of the boundary sites on the minimal density function for a given configuration
of classical particles. It is done in the isotropic case (that is, a
hypercubic Bravais lattice with equal hoppings in all directions) in \cite{FLU01}; the
extension to the anisotropic case is straightforward, but numerous details are
modified, and have been verified.

The results of the theorem have a number of implications
for the FK model.  It establishes that the segregation
principle holds in all dimensions (at $T=0$ and $U=\infty$)
illustrating the fact that the existence of periodic ground states requires a
subtle reduction in energy relative to the segregated phase as the
interaction strength is made finite.  Since the electronic wave functions
will be exponentially localized within $\Lambda$, the results shown
here can be extended to the case of finite interaction strength, as long
as $U$ is large enough.  At positive temperature it is so far impossible
to claim rigorous results, except the following weaker one: the electronic free
energy (for a {\it fixed}
configuration of classical particles) can be shown to be equal to the bulk free energy
plus a correction term that is proportional to the size of the boundary.
We expect that the coexistence of two phases occurs at finite
temperature for $d\ge 2$ as happens in the Ising model.

It may be instructive to consider the $U=\infty$
results for the FK model as a guide for
possible behavior in the H model. To do this, we must first find a way to interpolate between the
two models.
The simplest way is to consider an asymmetric-hopping H model
where the hopping for the spin-up and the spin-down particles is different.
Then the H model results when $t_\uparrow=t_\downarrow$ and the FK
model when $t_\downarrow=0$.  Our rigorous results only hold for
$t_\downarrow=0$.  When $t_\downarrow$ is increased the classical particles should still be
packed, due to the pressure of the electrons. When $t_\downarrow$ keeps increasing however,
the classical particles should be in a phase with density strictly less than 1.  The
central issue is whether the reduction of the down-spin kinetic energy
can be made large enough, so that the phase separation disappears at a critical
value of $t_\downarrow$. If this occurs for all electron densities,
there is no (saturated) ferromagnetism in the $U=\infty$ H model on the given lattice;
however, one has a saturated ferromagnetic ground state if the phase separation 
survives (note the $SU(2)$-imposed degeneracy of the ferromagnetic multiplet
will occur precisely at $t_\downarrow=t_\uparrow$).
It is well known that ferromagnetism depends strongly on the geometry of the
lattice~\cite{N63,L89}, so the occurrence of a critical
value of $t_\downarrow$ must also depend strongly on the geometry of the 
lattice.
We are unable to make any rigorous statements about ferromagnetism in the
H model here --- actually, we do not even know how to study the case
with nonzero, but small $t_\downarrow$.

A major question is what happens to the chessboard phase when doped away from
half filling? Consider the line $n_{\rm e} = n_{\rm c}$; the chessboard phase is
present when these densities are equal to $\frac12$, and segregation takes place
when they differ significantly from $\frac12$ (depending on $U$). It is not clear
what to expect for intermediate values. Two possible scenarios are (i) the
coexistence between chessboard and segregated phases  or (ii) the coexistence 
between other periodic and segregated phases.  Both scenarios could be of physical 
relevance to stripe physics.

We conclude this letter by a summary of our knowledge of the phase diagram
for zero temperature and $d\le 2$.
Recall that the particle-hole symmetry implies that the phase
diagram is symmetric under the transformation $(n_{\rm e}, n_{\rm c}) \mapsto (1-n_{\rm e},
1-n_{\rm c})$ \cite{KL86}. In the sequel we describe the situation for $n_{\rm e} +
n_{\rm c} \leq 1$, which is enough.

In two dimensions, the ground states are {\bf periodic}
\begin{itemize}
\item when $(n_{\rm e}, n_{\rm c}) = (\frac12,\frac12)$ (they are of the chessboard type). This
was proved for all $U$ in \cite{KL86};
\item when $n_{\rm c} = 1-n_{\rm e}$ (i.e.\ at half-filling), and $n_{\rm e} = \frac25,
\frac13, \frac14, \frac29, \frac15, \frac2{11}, \frac16$; also, $n_{\rm e} = \frac1{n^2 + (n+1)^2}$ with integers $n$.
This holds for $U$ large enough (depending on $n_{\rm e}$), and follows from 
\cite{GJL92,K94,K98,H98,H00};
\item when $n_{\rm c} = 1-n_{\rm e}$, and the electronic density $n_{\rm e}$ is a rational
number between $\frac13$ and $\frac25$. $U$ must be larger than a value that depends on the
denominator of $n_{\rm e}$ \cite{HK01}.
\end{itemize}

There is {\bf coexistence} of two periodic phases
\begin{itemize}
\item when $n_{\rm c} = 1-n_{\rm e}$ and $n_{\rm e} \in (\frac16,\frac2{11})
\cup (\frac15,\frac29) \cup (\frac29,\frac14)$, for $U$ large \cite{K98,H98}.
\end{itemize}

And the ground states display {\bf segregation}
\begin{itemize}
\item for $\frac{\rm const}U (1-n_{\rm c}) < n_{\rm e} < (1-\frac{\rm const}U) (1-n_{\rm
c})$; this is described here, and proved in \cite{FLU01}.
\end{itemize}

\begin{figure}
\epsfxsize=80mm
\centerline{\epsffile{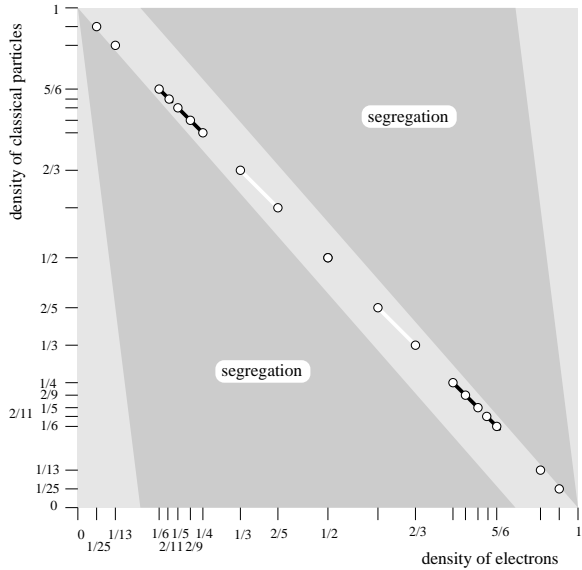}}
\caption{Schematic phase diagram of the 
rigorous results for the ground state of the 2D Falicov-Kimball model for large
$U$. The white dots and the white
lines represent periodic phases; the black lines are coexistences between
different periodic phases; and dark gray regions are segregated. There are no 
rigorous results for light gray domains.}
\label{FKrigresults}
\end{figure}

These results are illustrated in Fig.\ \ref{FKrigresults}.

The domain $\frac{\rm const}U
(1-n_{\rm c}) > n_{\rm e}$ should also
be segregated. The central band that includes the line $n_{\rm e} + n_{\rm c}
= 1$ should be the host of numerous periodic phases and various coexistences between
periodic phases and empty or full phases. This is supported by numerical
simulations in 2D \cite{2d_numerics}.

In one dimension, the ground states are {\bf periodic}
\begin{itemize}
\item
when $(n_{\rm e},n_{\rm c})=(\frac12,\frac12)$ \cite{KL86};
\item
when $n_{\rm c}=1-n_{\rm e}$ (half-filling) and $n_{\rm c}=p/q$ is a rational 
number (in an irreducible fraction).  The periodicity is $q$ for $U$ sufficiently large
(depending on $q$) \cite{L92}.
\end{itemize}

For finite $U$, there is numerical evidence for {\bf coexistence} of two
periodic phases and free electrons
\begin{itemize}
\item
when $n_{\rm c}=1-n_{\rm e}$, $p/q<n_{\rm c}<p^\prime/q^\prime$, and the 
periodic phases with period $q$ and $q^\prime$ are the only stable phases
within the above interval \cite{1d_numerics}.
\end{itemize}

And the ground states display {\bf segregation}
\begin{itemize} 
\item
when $n_{\rm e}\ne 1-n_{\rm c}$ and $U$ is sufficiently large \cite{L92,FLU01}.
\end{itemize}

The canonical phase diagram for small $U$ is even richer but our knowledge of it is very limited.

In conclusion, we have proved that the FK model is phase
separated whenever $N_{\rm e}<|\Lambda|$ and $U\rightarrow\infty$.  This shows how
strong correlations can lead to phase separation. 

J.K.F. acknowledges support from the Office of Naval Research under
grant N00014-99-1-0328. E.H.L. and D.U. acknowledge support from the
National Science Foundation under grant PHY-98-20650.  J.K.F. would
like to thank the hospitality of the Department of Physics at
Princeton University where the majority of this work was completed.
We thank S. Kivelson, R. Lema\'nski and M. Loss for helpful discussions.

\end{document}